\documentclass[a4paper]{jpconf}
\usepackage{graphicx}
\usepackage{epsfig}
\newcommand{\be}{\begin{eqnarray}}
\newcommand{\ee}{\end{eqnarray}}
\begin{document}
\title{Large extra dimensions and small black holes at the LHC }

\author{Marcus Bleicher and Piero Nicolini}

\address{Frankfurt Institute for Advanced Studies (FIAS), Johann Wolfgang Goethe-Universit\"{a}t, Ruth-Moufang-Str. 1, D-60438 Frankfurt am Main, Germany}

\address{Institut f\"{u}r Theoretische Physik, Johann Wolfgang Goethe-Universit\"{a}t, Max-von-Laue-Str. 1, D-60438 Frankfurt am Main, Germany}

\ead{bleicher@th.physik.uni-frankfurt.de, nicolini@th.physik.uni-frankfurt.de}

\begin{abstract}
We present an overview of the conjectured production of microscopic black holes as a consequence of high energy hadronic collisions at the Large Hadron Collider (LHC),  CERN (Geneva, Switzerland) from this year on. Provided the presence of large extra-dimensions, we analyze some possible scenarios that could turn out to be outstanding experimental discoveries. We also discuss some new models which have been recently proposed on the ground of quantum gravity arguments. The final comments are devoted to supposed potential risks connected with the formation of black holes in particle detectors.
\end{abstract}

\section{Introduction}
If we could look back at the scientific community in the 60's or in the 70's, we would observe something very unusual for the current time perspective: a clear separation between physicists working on the standard model of particle physics and those working on the theory of gravitation.
At that time the main stream of particle physics experiments were carried out at energies far below the supposed energy at which the fundamental interactions, gravity included, would have been unified. In other words gravity was simply assumed to be too
weak with respect to the other fundamental interactions and its long standing problem of lack of renormalization was somehow underestimated or at least considered relevant only at an unreachable energy scale.  Therefore gravity was kept classical and the major efforts were directed towards particle physics, calculating Feynman diagrams, cross sections and decay rates. On the other hand, General Relativity became the domain of mathematical physicists, who were mainly interested in the formal aspects of the theory, ``spoiling'' it from any physical content.
Against this background, in 1975 a paper by Steven Hawking ignited the interest for the reconciliation of particle physics and gravity \cite{Hawking:1974sw}. Indeed combining General Relativity and Quantum Field Theory, Hawking showed that black holes can emit thermal radiation like a black body.

After more than $30$ years of intensive research, the scenario is now
drastically different and we have at least two plausible candidate theories of
Quantum Gravity: Superstrings \cite{Schwarz:1982jn} and Loop Quantum Gravity\cite{Smolin:2004sx}. Both of them have great merits and still-open
problems, but at present both suffer from a basic limitation: the absence of
any support from experiments. For this reason they cannot be considered as
being more than theoretical speculations, even if they are both physically very
promising, aesthetically very attractive and mathematically elegant and well
defined.  The situation is even more puzzling if we think that the LHC, the world's largest and highest-energy particle accelerator, can reach just 14 TeV, an energy 16 orders of magnitude lower than the Planck scale, the energy at which, without further hypotheses, we think that gravity will be quantized and unified with the other fundamental interactions. On the side of cosmic ray observations, there are also nontrivial problems. On the one hand cosmic rays can reach energies over $10^{20}$ eV, on the other, their detection requires large detectors
with areas of at least many square kilometers to have a reasonable number of events per year. Given this point, there are only two ways out: on the experimental side to further improve the acceptance of cosmic ray detectors (as currently done e.g. with the AUGER observatory or ICECUBE) or to increase the energy of particle accelerators, while on the theoretical one can reformulate the scenario with further assumptions in order to predict the presence of quantum gravity effects at the current observable energy scales.  Since for now the ``experimental way out'' is slow, expensive and difficult we will analyze the ``theoretical'' one in the next section.

\section{The large extra-dimension scenario}
The new idea is that our universe might have more than the well known three spatial
dimensions, height, width and length, but up to seven additional space dimensions that
are usually unobservable. This hypothesis is currently considered to be the
unique viable solution to the long-standing hierarchy problem, namely the
presence of two fundamental scales in nature, the electroweak scale and the
Planck scale, separated by $16$ orders of
magnitude \cite{ArkaniHamed:1998rs,Antoniadis:1998ig,ArkaniHamed:1998nn,Gogberashvili:1999tb,Gogberashvili:1998iu,Gogberashvili:1998vx,Gogberashvili:1999ad,Kokorelis:2002qi,Randall:1999ee,Randall:1999vf}.
Conversely, in the
absence of any extra dimensions, Quantum Gravity maybe lost forever: it is
very likely that at the current rate of technological progress, mankind will
probably become extinct before any experimental evidence of Quantum Gravity
would become accessible. 
According to the Brane World scenario, the main point
in this potential resolution of the problem is that extra dimensions can be
assumed to be as large as around a fraction of a millimeter. To this purpose, we have to suppose that Standard
Model fields are spatially constrained on a 3-brane, actually living in a four dimensional sub-manifold of the higher
dimensional spacetime, while only gravity would be allowed to probe the transverse ``bulk'' dimensions. These additional dimensions are curled up to tiny ``doughnuts'' to make them hard to observe with a probe that is bigger than the extension of
the ``doughnut'' - i.e. the extra dimensions are compactified on a torus of radius $R$.
What makes these models especially interesting is the fact that the gravitational
interaction increases strongly (by factors on the order of $10^{32}$) on distances below the
extension of the extra dimensions. The strict way to show this is to start from the
Einstein-Hilbert action for $(k+4)$-dimensional gravity and to integrate out the additional
space like extra dimensions. However, there is an easier way to understand this.
Consider a particle of mass $M$ located in the bulk, a space time with $k$ + 3 dimensions. The
general solution of the Poisson's equation yields its potential as a function of the radial
distance $r$ from the source
\be
\phi(r)=\frac{1}{M_\star^{2+k}}\frac{M}{r^{1+k}}.
\label{extranewton}
\ee
Here we have introduced a new fundamental mass-scale $M_\star$ with a suitable exponent.
Remembering that the additional $k$ space-time dimensions are compactified on radii $R$,
then, at distances $r\gg R$, the extra dimensions should be 'hidden' and the potential Eq.
(\ref{extranewton}) should turn into the well known $1/r$ potential, however with a pre-factor including
the volume of the extra dimensions
\be
\phi(r)\rightarrow \frac{1}{M_\star^{2+k}}\frac{1}{R^k}\frac{M}{r}\equiv\frac{1}{M_P^{2}}\frac{M}{r}\quad {\mathrm as} \ \  r\gg R.
\label{newton}
\ee
In the limit of large distances, this should be identical to Newton's gravitational law,
yielding the relation
 \begin{equation}
M_\star^{(2+k)}=M_{P}^{2}/R^{k}
\end{equation}
Assuming that $M_\star$ has the right order of magnitude to be compatible with observed
physics, it can be seen that the volume of the extra dimensions suppresses the
fundamental scale and thus, explains the huge value of the Planck mass
\be
M_P\sim 1.22\times 10^{28}\mathrm{eV}.
\ee Today one
expects this new fundamental scale to be of the order of $M_\star\sim 1$ TeV, allowing sizes of
extra dimensions between 1/10 mm to 1/1000 fm for $k$ from 2 to 7. The case of one or two extra dimensions in the ADD scenario \cite{ArkaniHamed:1998rs} is actually ruled out by Cavendish like experiments, which confirm the validity of Newton's law in four dimensions  down to $1/10$ mm, while one would need an extension of at least $10$ km for $k=1$. More in detail, according to the latest results \cite{cavendish1,cavendish2}, the limit for the thickness of extra-dimension is  $R\le 44\ \mu{\mathrm m}$ which implies that the fundamental mass is
\be
M_\star\ge 3.2 \ \mathrm{TeV}.
\ee
This would mean that, according to the LHC schedule, we will be reaching the above energy in the first half of 2010.

\begin{figure}
\includegraphics[width=10cm]{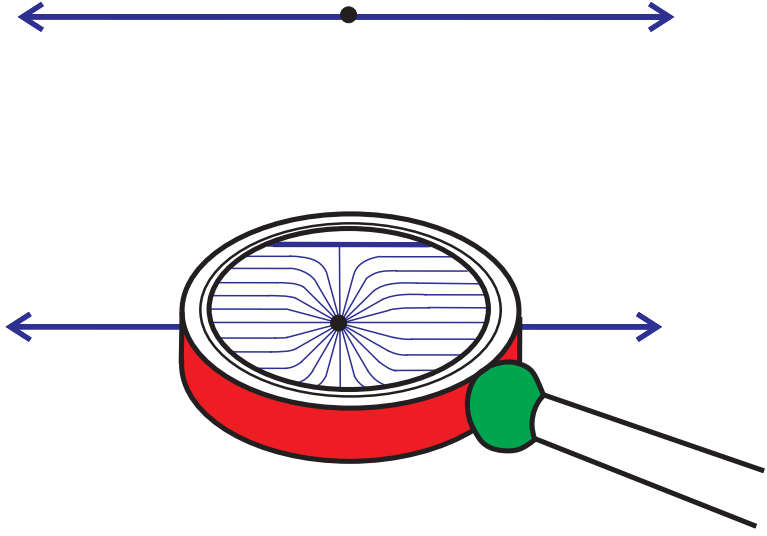}
\caption{\label{fluxlines} Flux lines emerging from a mass point. Top: Along one dimension in usual space time.
Bottom: Spread of the flux lines due to an additional extra dimension. Taken from \cite{Hossenfelder:2004af}.}
\end{figure}

The above arguments can be followed even clearer when one remembers that the
strength of interaction is proportional to the density of flux lines. Figure \ref{fluxlines} illustrates
this behaviour: while the top figure shows the setting with a mass point in the
center for one dimension with undiluted flux lines, the lower figure shows (in the
magnifying glass) that there are actually two dimensions and the field lines are
diluted. Mathematically speaking we find that in the usual three dimensional world
the gravitational potential falls-off like 1/distance, however if we have more dimension
(e.g. $k$ additional dimensions) to dilute the gravitational flux lines, the potential will
fall like 1/distance$^{k+1}$.
Why is the gravitational field then stronger at small distances? This becomes
clear, if we remember that the long distance behaviour of gravitation is fixed from daily
observation (the Newton's law). If we extrapolate towards smaller distances the additional
dilution of the field lines starts, therefore, to get the proper long distance behaviour back,
one has to start with more dense flux lines (meaning stronger gravitational interaction)
at small distances.
Thus, if Newton's law is modified only at small distances when extra dimensions are
introduced, it is obvious that the most promising experimental tests for the existence of extra dimensions may simply be
measurements of the Newtonian potential at small distances.

\section{Mini black hole production}

What is the role of black holes in this context? One of the most spectacular consequences of having a low fundamental scale $M_\star$ is the increase of both the
gravitational interaction strength and the probability of gravitational collapse, which is nothing but black hole formation. In fact, the increase in gravitational
strength is drastic enough that it might allow black hole production at the LHC as a result of high energy hadronic collisions, already within the next couple of months \cite{Argyres:1998qn,Dimopoulos:2001hw,Hsu:2002bd}.
However, how can one imagine the production of black holes in a particle accelerator? The nuclei of
two hydrogen atoms (the protons) are accelerated in opposite directions nearly to the
speed of light. When they collide, the subatomic constituents within the protons
interact. If the distance between these constituents (the partons) is small
enough - i.e. smaller than the corresponding Schwarzschild radius $R_H$ - and the energy is high
enough a micro black hole could be formed. The mass of such black holes will be of
the order of a TeV and therefore be much lighter than the cosmic
black holes we know up to now.
This can be strictly derived by using the higher dimensional Schwarzschild-metric
\cite{Myers:1986un}
\be
ds^2_{(k+4)}=-F(r)dt^2+ F^{-1}(r)dr^2+r^2d\Omega^2_{3+k},
\ee
where $d\Omega^2_{3+k}$ is the surface element of the $3 + k$ - dimensional unit sphere and
\be
F(r)=1-\left(\frac{R_H}{r}\right)^{k+1}.
\label{eq7}\ee
The constant $R_H$ can be obtained by requiring that the solution reproduces the
Newtonian limit for $r\gg R$. This means the derivative of Eq. (\ref{eq7}) has to yield the Newtonian
potential in $(3+k)$ dimensions:
\be
\frac{1}{2}\frac{dF(r)}{dr}=\frac{k+1}{2}\left(\frac{R_H}{r}\right)^{k+1}\frac{1}{r}=\frac{1}{M_\star^{k+2}}\frac{M}{r^{k+2}}.
\ee
So we have
\be
F(r)=1-\frac{2}{k+1}\frac{1}{M_\star^{k+2}}\frac{M}{r^{k+1}},
\ee
and $R_H$ is
\be
R_H^{k+1}=\frac{2}{k+1}\left(\frac{1}{M_\star}\right)^{k+1}\frac{M}{M_\star}.
\ee
For $M_\star\sim 1$ TeV this radius is $\sim 10^{-4}$ fm. Thus, at high enough energies it might be
possible to bring particles closer together than their horizon, which will force them to
form a black hole.

\begin{figure}
\epsfig{figure=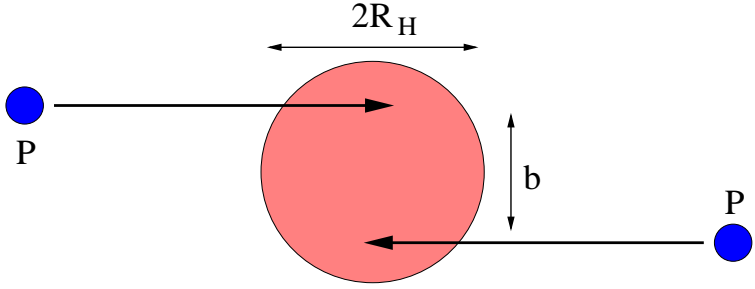,width=10cm }  
\caption{At very high energies, the partons, $p$,
can come closer than the Schwarzschild radius, $R_H$, associated with their energy. If the
impact parameter, $b$, is sufficiently small, such a collision will inevitably generate
a black hole.}
\label{bumm}
\end{figure}

Let us now calculate the possible production rate of black holes explicitly. That is,
we consider two elementary particles, approaching each other with a very high kinetic
energy in the c.o.m. system slightly above the new fundamental scale $M_\star\sim 1$ TeV, as
depicted in figure \ref{bumm}. Since the black hole is not an ordinary particle of the Standard Model and its correct quantum theoretical treatment is unknown, it is treated as a metastable
state, which is produced and decays according to the semi classical formalism of
black hole physics.
To compute the formation probability, we approximate the cross section of the black
holes by their classical geometric area
\be
\sigma(M)\approx \pi R_H^2\Theta(M-M_\star),
\ee
with $\Theta$ being the Heaviside function. Setting $M_\star\sim 1$ TeV and $k = 2$ one finds
$\sigma\sim 400$ pb. Using the geometrical cross section formula, it is now possible to compute
the differential cross section  $d\sigma/dM$ which will tell us how many black holes will be
formed with a certain mass M at a c.o.m. energy vs the probability that two colliding
particles will form a black hole of mass M in a proton-proton collision at the LHC (at
a center of mass energy of 14 TeV) involves the parton distribution functions. These
functions, $f_A(x,\hat{s})$, parametrise the probability of finding a constituent $A$ of the proton
(quark or gluon) with a momentum fraction $x$ of the total energy of the proton. 
Here, $\sqrt{\hat{s}}$ is the c.o.m. energy of the parton-parton
collision.
The differential cross section is then given by summation over all possible parton
interactions and integration over the momentum fractions $0\le x_1 \le 1$ and $0\le x_2 \le 1$,
where the kinematic relation $x_1x_2s = \sqrt{\hat s} = M^2$ has to be fulfilled. This yields
\be
\frac{d\sigma}{dM}=\sum_{A_1 B_1}\int_0^1 dx_1\frac{2\sqrt{\hat {s}}}{x_1 s}f_a(x_1, \hat{s})f_B(x_2, \hat{s})\sigma (M,k).
\label{diffcross}
\ee
The particle distribution functions for $f_A$ and $f_B$ are tabulated e.g. in the CTEQ-tables\footnote{The parton distribution function can be downloaded from this web page:
http://www.phys.psu.edu~cteq/}. One might think that the mass distribution diverges when $x_1 = 0$, however, this
limit does not fulfill the kinematic constrained $x_1x_2 = M^2/s$, because it would result in
$x_2 > 1$ for finite black hole masses $M = M_\star$ . It is now straightforward to compute the total cross section and number by
integration over Eq. (\ref{diffcross}) This also allows us to estimate the total number of black
holes, $N_{\mathrm{BH}}$, that would be created at the LHC per year. It is
\be
N_{\mathrm{BH}}/\mathrm {year} = \sigma(pp \rightarrow \mathrm{BH})L,
\ee
where we insert the estimated luminosity for the LHC, $L = 10^{33}\ \mathrm{cm}^{-2}\mathrm{s}^{-1}$. This yields at
a c.o.m. energy of $\sqrt{s} = 14$ TeV the total number of 10 million black holes per year! This
means, about one black holes per second would be created. Even if this is only an optimistic upper 
estimate of the black hole formation rate, the importance of
this process led to a high number of publications on the topic of TeV-mass black holes
at colliders both in scientific journals \cite{Banks:1999gd,Bleicher:2001kh,Hossenfelder:2001dn,Giddings:2001bu,Cavaglia:2003qk,Cavaglia:2004jw,Tanaka:2004xb,Harris:2004xt,Godang:2004bf,Gingrich:2009hj,Aref'eva:2009ng,Gal'tsov:2009zi} and in the popular press\footnote{``Physicists Strive to Build A Black Hole'', New York Times, September 11, 2001.}.
It should be noted that numerical simulation and visualisation packages for the production and decay of black holes in
colliders and from cosmic rays are available. These packages are currently used to explore
potential signatures for black hole production and go by the names CHARYBDIS \cite{Harris}, CHARYBDIS2 \cite{Frost:2009cf}, TRUENOIR\cite{Dimopoulos:2001en},
GROKE \cite{groke1,Ahn:2005bi}, CATFISH \cite{catfish} and, more recently, QBH \cite{Gingrich:2009hj,Gingrich:2009da}  and  BlackMax \cite{Dai:2007ki} which employs a richer set of external parameters and allows a certain flexibility among different theoretical models.

\section{Mini black hole evaporation}
Hawking's revolutionary theoretical discovery about the possibility for black hole to emit thermal radiation like a black body now enters the scene, disclosing new scenarios. Indeed it is very likely that black holes are fated to provide the final answer about our knowledge of Quantum Gravity \cite{Feng:2001ib,Cavaglia:2002si,Stoecker:2006yz,Casadio:2001wh,Giddings:2001ih,Stenmark:2002yb,Chamblin:2004zg,Rizzo:2006uz}  and close the logical path which they started in 1975, igniting the initial interest in this area.
According to theory, black holes would have a temperature
\be
T_H=\frac{1}{4\pi}\left.\frac{dF(r)}{dr}\right|_{r=r_H},
\ee
and their decay would lead to the sudden emission of a huge quantity of particles independently of their spin, charge, quantum numbers or interaction properties, as long as their rest mass is smaller than the black hole temperature \cite{Casanova:2005id,Kanti:2004nr}.
Therefore mini black holes may be excellent intermediate states for long-sought
but yet undiscovered particles, like the Higgs boson or supersymmetric particles, possibly with cross sections
much larger than for the direct processes \cite{Erkoca:2009kg}. We could also have key information about the very early universe, as well as solving some basic questions, whose answers are too often taken
for granted. 

First, do (mini) black holes exist? From velocity measurements for the whirlpool of hot gas surrounding it,
astronomers have found convincing evidence for the existence of a supermassive black hole in the center of the
giant elliptical galaxy $\mathrm{M}87$. Astronomers have also
detected radio emission coming from within 30 million kilometers of the dark object SGR A*, thought to be a
colossal black hole, that lies at the center of the Milky Way \cite{Ghez:2003hb}.
Previously, X-ray emission from the binary star system Cygnus X-1 convinced many astronomers that it contains a
black hole, which is supported by more precise measurements which have recently become available. To this purpose
we also have to mention V404 Cygni, one of the most evident cases for a stellar black hole: observations at X-ray and
optical wavelengths have shown that this is a binary system in which a late yellowish G star or maybe an early
orange-red K star revolves, every 6.47 days, around a compact companion with a diameter of around $60-90$ km and a
probable mass of 8 to 15 solar masses, well above the mass limit at which a collapsed star must become a black
hole \cite{Corbel:2008yw}. In spite of these observations, however, there are still some
ranges of mass in which the existence of black holes is unclear, in particular of black holes of less than $3$
solar masses. The relevance of these objects is connected with the possibility of observing the Hawking radiation.
Indeed, what we know for sure is that for astrophysical black holes the Hawking radiation is negligible because
their temperatures can be at most some tens of nK, far below $T_{\mathrm{CMB}}\sim 2.7$ K, the temperature of
Cosmic Microwave Background (CMB) radiation. On the other hand tidal effects are significant in the case of mini
black holes that could be very hot and very bright if their mass is sufficiently small \cite{Page:1976wx}.

As second point, we might be able to conclude that (mini) black holes really can evaporate. Indeed for the above
reasons the detection of these objects is the unique direct way to have experimental evidence of the
Hawking conjecture, one of the most important predictions of Quantum Field Theory in Curved Spacetime and of the
associated semiclassical gravity.

Thirdly, we could find out what the fate of a radiating black hole is. If
mini black holes can be created in high energy particle collisions, the black
holes produced will pass through a number of phases before completing their
evaporation.  After a loss during the first two phases of their ``hair'' (i.e.
the associated long-range fields) and of their angular momentum (the
``balding'' and spin-down phases respectively), the picture of the evaporation
will be described by the Schwarzschild phase, in which the resulting
spherically symmetric black hole loses energy by the emission of Hawking
radiation, with a gradual decrease in its mass and an increase in its
temperature. 
Since the Schwarzschild geometry has a curvature singularity at
the origin, there would be a divergent behavior of the Hawking temperature if
the black hole were to shrink to the origin as a result of losing mass by
thermal emission. However, we do not expect that this divergent behavior will,
in fact, take place since in the vicinity of the origin, the evaporating black
hole will be dramatically disturbed by strong quantum gravitational
fluctuations of the spacetime manifold. In other words the black hole will
undergo a Plank phase of the evaporation during which a theory of Quantum
Gravity must be used. Observations of the final stages of black hole
evaporation could provide the profile of the temperature as a function of the
mass of the black hole and hence let us pick out, for the first time, the
correct quantum gravitational theory.

\section{New developments}

At this point, it is clear that the the semiclassical approximation given by Quantum Field Theory in
Curved Space, turns out to be inadequate when the black hole undergoes the most important phase of the evaporation, the final phase. To this purpose there exists the possibility of improving semiclassical gravity with some arguments coming from String Theory. If we believe in the Brane World scenario and the presence of a $3$-brane where all massive objects live, we have to recall that it has been shown that open string end point coordinates do not commute on branes, giving rise to the so called Noncommutative (NC) Geometry \cite{Seiberg:1999vs}
\be
\left[{\bf x}^\mu, {\bf x}^\nu\right]=i\Theta^{\mu\nu},
\ee
with $x^\mu, x^\nu$ being the 4-coordinates and $\Theta$ being the non-commutativity parameter.
Even if we are not fully quantizing gravity, in this approach we obtain an induced model of quantum spacetime. The quantum nature of the manifold can be observed in the consequent loss of resolution coming from the uncertainty relation
\be
\Delta x^\mu \Delta x^\nu\ge \theta
\ee
where $\sqrt\theta$ is the universal quantum of length. Since there is  no  way to determine the
exact position of any physical object, we cannot longer speak of pointlike objects because there are no points on such a brane.
As a result it can be shown that the conventional Dirac delta distributions $\delta({\bf x})$ switches to a minimal width  Gaussian distributions $\rho_\theta ({\bf x})$, where the minimal width is $\sqrt{\theta}\sim 1/M_\star$. Therefore if one solves the Einstein equations with an energy momentum tensor written in terms of minimal width  Gaussian distributions, one obtains a string improved semiclassical approximation, namely gravity + NC geometry \cite{Nicolini05,Nicolini05bis,Nicolini06,Rizzo06,Spallucci06,Ansoldi07,Spallucci:2008ez,Nicolini09,Arraut:2009an,Nicolini:2009gw,Nicolini:2009dr,Banerjee:2009xx,Modesto:2009qc,Batic:2010vm}. Looking for spherically symmetric black hole solutions, one assumes a metric of the form
\begin{equation}
ds^2_{(d+1)}= -F(r)\, dt^2 + F^{-1}(r)\, dr^2 + r^{2} d\Omega^2_{d-1}, \label{extrads}
\end{equation}
where now $d=3+k$ is the total number of spatial dimensions. Plugging the above line element in
\begin{equation}
R_M^N=\frac{1}{M_\star^{d-1}}\left(8\pi\ T_M^N-\frac{8\pi}{d-1}\ \delta_M^N\ T\right),
\end{equation}
where $T$ is the trace of the energy momentum tensor, $T=T_M^M$, one finds the noncommutative black hole solution
\begin{equation}
F(r)= 1 -\frac{1}{M_\star^{d-1}}\frac{ 2M }{r^{d-2} \Gamma(d/2)}  \,\, \gamma\left(\, \frac{d}{2}\ ,
\frac{r^2}{4\theta}\,\right)
\end{equation}
where
\begin{equation}
 \gamma\left(\, d/2\ , r^2/4\theta\,\right)\equiv
  \int_0^{r^2/4\theta} \frac{dt}{t}\, \, t^{d/2} \, e^{-t},
\end{equation}
is the incomplete gamma function and
\begin{eqnarray}
&\Gamma\left(d/2\right)= \left(\frac{d}{2}-1\right)!\hspace{10mm}  d\,\, even&\\
&\Gamma\left(d/2\right)= \sqrt{\pi}\ \frac{(d-2)!!}{2^{(d-1)/2}}\hspace{10mm}  d\,\, odd.&
\end{eqnarray}
To start commenting the above result we check first of all that at large distances with respect to the quantum of length, stringy effects become negligible and the conventional Schwarzschild solution is reproduced. Indeed for $\sqrt{\theta}/r\to 0$ the incomplete gamma functions $\gamma\left(\, \frac{d}{2}\ ,
\frac{r^2}{4\theta}\,\right)$ match the value of the Euler functions $\Gamma(d/2)$ and the Schwarzschild solution is obtained.
The novelty of the solution lies in the following new properties
\begin{itemize}
\item the curvature singularity is replaced by a de Sitter core that makes the solution regular and accounts for the quantun gravitational fluctuation, namely for $r\ll\sqrt\theta$
 \begin{equation}
F(r)= 1 - \frac{1}{d\, \, M_\star^{d-1}}\frac{4M }{ 2^{d-1}\pi^{(d-2)/2 }\ \theta^{d/2}}\, \, r^2 +{\cal O}\left(\,
r^4\,\right). \label{desit}
\end{equation}
\item the horizon equation  $F(r)=0$ has three possibilities depending on the value of the mass $M$,
namely two horizons $r_+$ and $r_-$, one single degenerate horizon $r_0$ and no horizon. 
\end{itemize}

 \begin{figure}[ht!]
 \begin{center}
\includegraphics[width=10cm,angle=270]{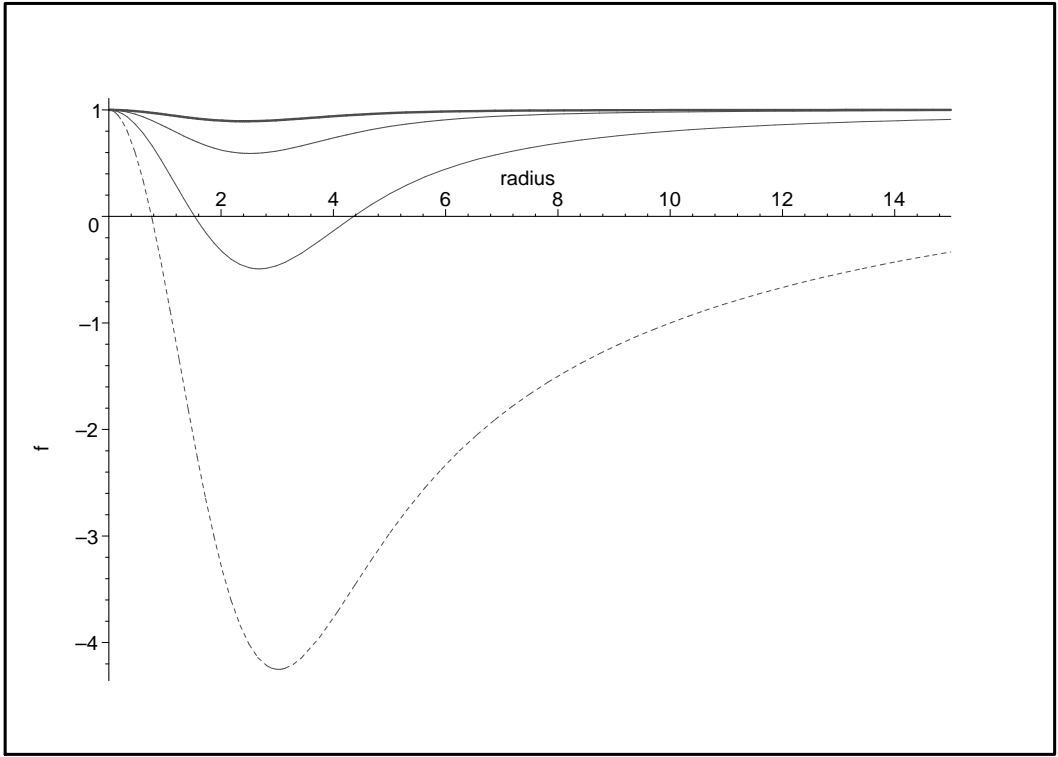}
\caption{\label{mg00} \ {\it The noncommutative Schwarzschild solution.} The function $F$ is
plotted versus $r/\sqrt{\theta}$, for $M=10\, M_\star $. We can observe that the curves rise with $d$ and the outer
horizon $r_+$ decreases until $d=4$. For $d\ge 5$ no black hole can be formed: the mass $M$ is so light that it
cannot provide a significant gravitational disturbance, since $F\simeq 1$. }
\end{center}
\end{figure}

\begin{figure}[ht!]
\begin{center}
\includegraphics[width=10cm,angle=270]{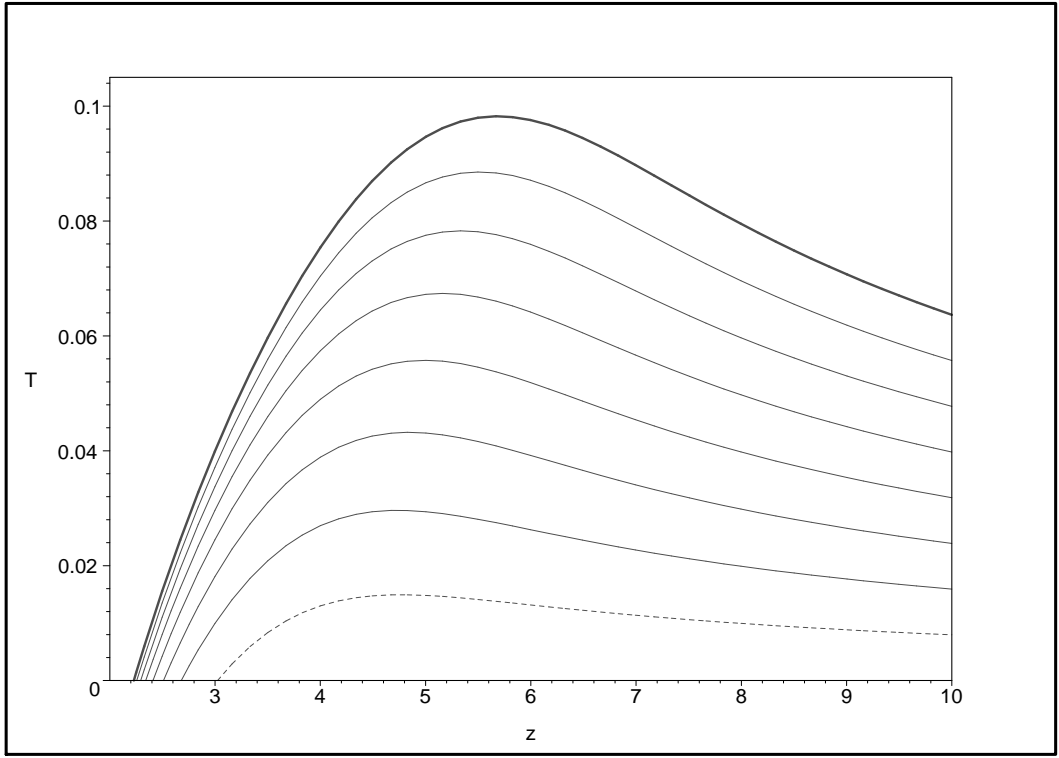}
\caption{ \label{bhtemp_max} \textit{The noncommutative extra-dimensional Schwarzschild solution.} The temperature
$T_H$ as a function of $z=r_+/\sqrt{\theta}$, for various value of $d$. The lowest curve corresponds to the case $d=3$, while, from
the bottom to top on the right hand side of the figure, the solid lines correspond to $d=4$ to $10$.  The
temperatures increase with $d$, while the SCRAM phase leads to smaller remnant for higher $d$.}
\end{center}
\end{figure}

\begin{table}
\caption{\label{ta1}Black hole maximal temperatures for different $d$. See figure (\ref{bhtemp_max}).}
\begin{center}
\begin{tabular}{cccccccccc}
\br
d
& & 3  & 4  & 5  & 6  & 7  & 8  & 9  & 10 \\
\mr
$T_H^{max}$ (GeV)
& & $18\times 10^{16}$  & 30  & 43  & 56  & 67  & 78  & 89  & 98 \\
$T_H^{max}$ ($10^{15} K$) & & $.21\times 10^{16}$  & .35  & .50  & .65  & .78  & .91  & 1.0
& $1.1$ \\
\br
\end{tabular}
\end{center}
\end{table}

\begin{table}
\caption{\label{ta2}Remnant masses and radii for different values of $d$, keeping $M_\star\sim 1/\sqrt{\theta}$}
\begin{center}
\begin{tabular}{ccccccccc}
\br
& 3 &  4 &  5 &  6 &  7 &  8 &  9 &  10 \\
\mr
$M_0$ (TeV)
 & 2.3 $\times 10^{16} $  & $6.7$  & $24$  & $94$  & $3.8\times 10^2$
 & $1.6\times 10^3$  & $7.3\times 10^3$  & $3.4\times 10^4$ \\
$r_0$ ($10^{-4}$ fm)  & $4.88 \times 10^{-16} $   & $5.29$  & $4.95$  & $4.75$  & $4.62$  &
$4.52$  & $4.46$  & $4.40$ \\
\br
\end{tabular}
\end{center}
\end{table}

The plot in figure \ref{mg00} shows that the outer horizon radius $r_+$ decreases as $d$ increases for a given mass $M$. The minimum of each
curve also provides the value of the extremal radius $r_0$ which decreases with $d$. 

The next step is about the thermodynamical analysis of the noncommutative black hole.  Here the temperature is
\begin{equation}
 T_H = \frac{d-2}{4 \pi r_+}\,\left[\, 1 -\frac{ r_+}{d-2}
 \frac{\gamma^\prime\left(\, d/2\ ,r_ +^2 /4\theta\,\right)}
 {\gamma\left(\, d/2\ , r_ +  ^2 /4\theta \,\right)}\,\right].
 \label{extratemp}
\end{equation}
From the above formula one obtains again a solution that behaves like the Schwarzschild solution at large distances, thus, reproducing the Hawking result $T_{H\,{\mathrm comm}}=(d-2)/4\pi r_+$.
 Noncommutative effects appear as $r_+$ approaches $\sim 6\sqrt\theta$, depending on $d$: the manifold is more sensitive to noncommutativity for higher dimensions.
Then, from figure \ref{bhtemp_max}, we conclude that  after an initial Schwarzschild phase, the black hole
undergoes a SCRAM\footnote{The term SCRAM, probably the
backronym for ``Safety Control Rod Axe Man'' or ``Super Critical
Reactor Axe Man'', refers to an emergency shutdown of a
thermonuclear reactor. The term has been extended to cover
shutdowns of other complex operations or systems in an unstable
state, but it also has a ``standard'' meaning ``go away quickly'',
in particular when we address children or animals. } phase, namely a cooling down towards a zero temperature extremal black hole remnant. This is the equivalent on the thermodynamics side of the absence of any curvature singularity: no divergent Planck phase occures. One can see also from table \ref{ta1} that the
peak of the temperature increases with $d$, but even in the case $d=10$ it is about $98$ GeV $\left(\, \simeq
10^{15}\right.$ K$\left.\right)$ which is much lower than $M_\star$. 

Remnant masses can be calculated and  their values together with remnant radii are summarized in table~\ref{ta2}. Since remnant radii are $\sim 4-5\times10^{-4}$ fm, i.e. smaller than the size of extra-dimensions, these remnants can be considered, to a good approximation, as totally submerged in a $d+1$ dimensional isotropic spacetime, for any $d=3-10$, confirming the validity of the spherically symmetric noncommutative solution. As a further
consequence, estimating the black hole production cross section as the area of the event horizon
$\sigma \sim \pi r_+^2$, we find, for every $d$, an encouraging lower bound $\sigma\sim 10$ nb, almost two orders of magnitude
larger than the conventional values. On the other hand, for $d\ge 5$ the remnant is too heavy to be produced at
the LHC and could be only detected in Ultra-High-Energy cosmic rays \cite{Cavaglia:2002si}.

As a final comment about the String Theory induced noncommutative black holes, we have to say that, starting from different approaches, like Loop Quantum Gravity and Asymptotically Safe Gravity, there exist other black hole solutions which have a SCRAM phase, a regular geometry at the origin and one, two or no horizon \cite{Modesto:2006mx,Modesto:2008im,Modesto:2009ve,Hossenfelder:2009fc,Bonanno:2000ep,Bonanno:2006eu,Burschil:2009va,Scardigli:1999jh,AmelinoCamelia:2005ik,Husain:2008qx,Jizba:2009qf}. We can conclude that the depicted behavior for noncommutative black holes is universal, once Quantum Gravity is invoked, i.e. independent of the formulation.

\section{Final remarks}

The production of microscopic black holes at the LHC has recently raised a hot debate about the safety of high energy physics experiment \cite{fox}.
If, for some reasons, black holes do not evaporate, it was speculated that they may be potentially dangerous, since classically they may start to accrete indefinitely \cite{Koch:2008qq}. 
However, accurate studies have shown that the black hole lifetimes are extremely small. Even in the most pessimistic scenario according to which the black hole has a SCRAM phase and therefore can live longer than expected in conventional theories the lifetimes cannot exceed $\sim 10^{-16}$ s, taking into account all kinds of (bulk/brane) emissions  and number of spatial dimensions \cite{Casadio08}.
Also the destiny of eventual stable black hole remnants is not a source of concern.  
The probability that a black hole remnant starts
to grow inside the Earth is proportional to its volume and therefore negligible \cite{Hossenfelder:2003dy,Hossenfelder:2005ku,Koch:2005ks,Bleicher:2007hw,Koch:2007zza,Koch:2007um}.
To this purpose, accurate studies performed in \cite{Giddings:2008gr,Giddings:2008pi,Koch:2008qq,Casadio:2009ri,Casadio:2009sz,Gingrich:2010ce}, addressing the specific case of micro black holes in particle detectors, have excluded any threat from these kind of experiments at the LHC.

\ack
This work is supported by the Helmholtz International Center for FAIR within the
framework of the LOEWE program (Landesoffensive zur EntwicklungWissenschaftlich-\"{O}konomischer
Exzellenz) launched by the State of Hesse. We would like to thank Michelangelo Mangano for valuable comments and Martin Sprenger for discussions.

\end{document}